\documentstyle[12pt]{article}
\begin{document}
\baselineskip 18pt
\begin{center}
{\Large Spontaneous Leptogenesis}\\
\vskip .3in
{\large Utpal Sarkar  }\\[1cm]

{\large Theory Group }\\ {\large  Physical Research Laboratory}\\
{\large Ahmedabad-380009, India}\\

\end{center}

\vskip  .75in  
\begin{abstract}  
\baselineskip  18pt 

I  propose  a  new  mechanism  for   baryogenesis,  in  which  the
out-of-equilibrium   condition   is  not   neccessary.  When  the
electroweak   symmetry  is  broken   spontaneously,   left-handed
neutrinos  may  get  Majorana  masses  containing  $CP-$violating
phases.  This induces a lepton asymmetry  spontaneously, which is
then  converted  to a baryon  asymmetry  in the  presence  of the
sphaleron field.

\end{abstract}

\newpage
\baselineskip 18pt

The  baryon   asymmetry  of  the  universe  has  been   discussed
extensively   in   the   literature   \cite{sak,kolb}.  All   the
mechanisms  require to fulfill the three  conditions  proposed by
Sakharov   \cite{sak}.  It  requires   baryon  number   violating
interaction,  which  also  violate  $C$ and $CP$  symmetry.  This
interaction  should  be  slower  than the  expansion  rate of the
universe, {\it i.e.,} satisfy the out-of-equilibrium condition.

In  the   present   article  I  propose  a  new   mechanism   for
baryogenesis,  where  the  out-of-equilibrium  condition  is  not
required.  Lepton  number  and  $CP$  violation  comes  from  the
Majorana  mass  matrix  of the  left-handed  neutrinos.  As  time
evolves these massive Majorana  neutrinos  remains in equilibrium
and hence satisfy Boltzmann statistics.  But due to the rephasing
invariant  $CP-$violating  creation  phases in the Majorana  mass
matrix the  population  of the fields  carrying  lepton  number 1
becomes  different  from the population of the fields  carrying a
lepton   number  -1.  A  lepton   asymmetric   universe  is  thus
spontaneously   created   when  the   electroweak   symmetry   is
spontaneously  broken  \footnote{We  studied the possibility of a
lepton  asymmetric  universe  for the  heavy  Majorana  neutrinos
recently   and  here  a  similar   technique   will  be   adopted
\cite{pas1,pas2}.}.  This lepton asymmetry is then converted to a
baryon  asymmetry  in  the  presence  of the  sphalerons  through
anomalous baryon number violating interaction.

I work in the context of the  standard  model.  For the  neutrino
mass I donot  assume any  particular  mechanism.  In the  see-saw
models  \cite{see-saw}, there exists right handed neutrinos whose
Majorana   masses  breaks  the  lepton  number.  Whereas  in  the
Zee-type  models  \cite{zee},  a $SU(2)_L$  singlet  higgs scalar
breaks  the lepton  number.  We donot  subscribe  to any of these
specific  models for this mechanism to work.  We only assume that
lepton number is broken before the electroweak symmetry breaking.
As a result when the higgs doublet of the standard model acquires
a vacuum expectation value ($vev$), the left-handed neutrinos get
Majorana  masses  (since  there is no  symmetry  to  prevent  it)
through see-saw mechanism or through radiative corrections.  Once
the  neutrinos  acquire  masses  the  $CP$  symmetry  can also be
violated  spontaneously.  Similar to the $CP-$violating  phase in
the CKM  matrix,  there  can be new  $CP-$violating  phase in the
leptonic sector.

The Majorana mass term of the left-handed neutrinos ($\nu_i$, where
i = 1,2,3 corresponds to three generations), which can be written as,
\begin{equation}
{\cal L} = \sum_{ij} \tilde{m}_{ij} \overline{(\tilde{\nu_i})^c} 
\tilde{\nu_j} 
\end{equation}
can have its  origin in the  see-saw  mechanism  or the  Zee-type
models.  But for the present purpose we start with this effective
Majorana  mass term in the  lagrangian.  This mass matrix can, in
general, be complex and can have  $CP-$violating  complex  phase.
But the  fields  $\tilde{\nu_i}$  can now be  rotated  to a basis
$\nu_i = {\cal N}_{ij}  \tilde{\nu_j}$, where the new mass matrix
${\cal M}={\rm diag} \:\:  (m_1, m_2, m_3)$ is real and diagonal,
\begin{equation}
{\cal L} = \sum_{i} {m}_{i} \overline{({\nu_i})^c} {\nu_i} .
\end{equation}
However, this will introduce  $CP-$violating  phase in the charge
current and higgs  interactions of the neutrinos with the charged
leptons.  So  although  the tree level  Majorana  mass  matrix is
diagonal  and  real  in  this  basis,  there  will  be  radiative
corrections  as  given  in  figure  1,  which  will  then  induce
off-diagonal terms (in general complex) in the mass matrix.
\begin{figure}[hbt]
\mbox{}
\vskip 1.75in\relax\noindent\hskip -.425in\relax
\includegraphics{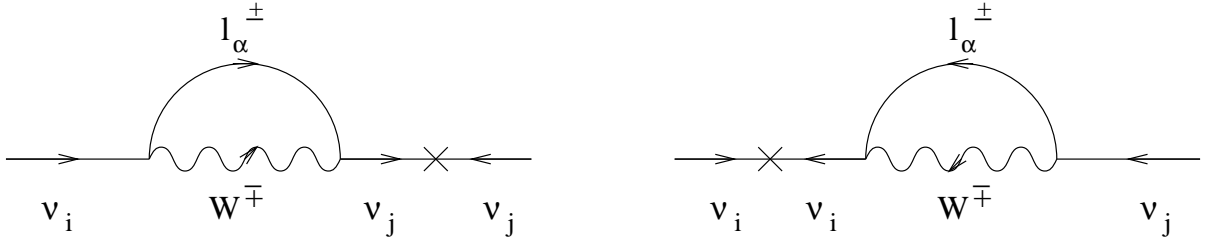} \vskip .2in
\caption{ One loop graph contributing to the Majorana masses of
the left-handed neutrinos.}
\end{figure}

For the sake of simplicity and to make the physics of the promlem
clear I consider only two  generations.  Since we want to discuss
the  question of  $CP-$violation,  we  distinguish  the  neutrino
fields $|\nu>$ and the  anti-neutrino  fields  $|\nu^c>$ to start
with and then show that there is only one physical Majorana field
at the end. From the charge current interactions of the fields 
$|\nu>$ and  $|\nu^c>$ with the charged leptons 
\begin{equation}
{\cal L} = g \overline{\nu} \gamma_\mu l^- W^{\mu +} + 
g \overline{\nu^c} \gamma_\mu l^+ W^{\mu -}
\end{equation}
it is possible to 
assign  lepton numbers 1 to $|\nu>$ and -1 to $|\nu^c>$ respectively.

To get the physical Majorana eigenstates of the problem we start
with the effective Hamiltonian of this model given in terms of the tree
level real masses $m_1$ and $m_2$ and the one loop corrections 
arising from the self-energy type diagrams of figure 1 \footnote{
There will be similar diagrams with the higgs scalars instead of the
$W^\pm$, but those diagrams will be suppressed by the Yukawa couplings
of the leptons.}. The
effective Hamiltonian in the  basis  \hbox{($|\nu_1^c>  \quad
|\nu_2^c> \quad |\nu_1> \quad |\nu_2>$)} now reads
\begin{equation}
{\cal H} = \pmatrix{0 & 0 & m_1 + \hat{m_1} & m \cr
0 & 0 & m & m_2 + \hat{m_2} \cr m_1 + \hat{m_1} & \tilde{m} &0 & 0 \cr 
\tilde{m} & m_2 + \hat{m_2} &0 & 0}
\end{equation}
where,
\begin{eqnarray}
{m} &=& g^2 \left[m_1 \sum_\alpha V_{\alpha i}^\ast V_{\alpha j}
+ m_2 \sum_\alpha V_{\alpha i} V_{\alpha j}^\ast \right]
(g^d_{\alpha i j} - {i \over 2} g^a_{\alpha i j} ) \\
\tilde{m} &=& g^2 \left[m_1 \sum_\alpha V_{\alpha i} V_{\alpha j}^\ast
+ m_2 \sum_\alpha V_{\alpha i}^\ast V_{\alpha j} \right]
(g^d_{\alpha i j} - {i \over 2} g^a_{\alpha i j} ) \\
\hat{m_i} &=& g^2 \left[2 m_i \sum_\alpha V_{\alpha i} V_{\alpha i}^\ast
\right]  (g^d_{\alpha i i} - {i \over 2} g^a_{\alpha i i} ) 
\end{eqnarray}
where $g$ is the $SU(2)_L$  gauge coupling  constant,  $V_{\alpha
i}$ is the  mixing  matrix in the  charge  current  interactions
in the basis where both the neutrino and charge lepton mass matrices
are diagonal. The  index  $\alpha$  corresponds  to the  charged  lepton
exchanged  in the diagram.  $V_{\alpha  i}$ is similar to the CKM
matrix in the quark  sector,  except that now due to the Majorana
nature of the neutrinos there is one $CP-$violating phase even in
the two generation  case.  The dispersive  part of
the loop integral  $g^d_{\alpha i j}$ can be absorbed in the wave
function and the mass  renormalization.  The  absorbtive  part of
the loop integral  $g^a_{\alpha i j}$ is  nonvanishing as long as
the momentum of the external neutrino fields are large and satisfy
the condition, $p m_W > m_l^2$. We are interested in the range of 
temperatures 250 GeV (when a higgs acquire a $vev$) to about 50 GeV
(when the sphaleron transitions become ineffective).  
In this limit the absorbtive  part of the integral is
given by, $$ g^a_{\alpha i j} = {1 \over 16 \pi} .$$

The Hamiltonian of equation(4) can be solved exactly 
\cite{pas2}, but the expressions
become somewhat involved. On the other hand without loss of 
generality it is possible to demonstrate the basic idea of the
problem assuming a mass hierarchy
$m_1 > m_2 >> m, \tilde{m}, \hat{m_i}$. In this limit one can use
perturbation theory to get the physical eigenstates,

\begin{eqnarray}
|\nu^{\rm phys}_1> & = & \frac{1}{\sqrt{\cal N}} 
(|\nu_1> + \alpha_2 |\nu_2> + |\nu_1^c>  + \alpha_1 |\nu_2^c>) \nonumber \\
|{\nu_1^{\rm phys}}^\prime> & = & \frac{1}{\sqrt{\cal N}} 
(|\nu_1> + \alpha_2 |\nu_2> - |\nu_1^c> - \alpha_1 |\nu_2^c>) \nonumber \\
|\nu^{\rm phys}_2> & = & \frac{1}{\sqrt{\cal N}} 
(|\nu_2> - \alpha_1 |\nu_1> + |\nu_2^c> - \alpha_2 |\nu_1^c>) \nonumber \\
|{\nu_2^{\rm phys}}^\prime> & = & \frac{1}{\sqrt{\cal N}} 
(|\nu_2> - \alpha_1 |\nu_1> - |\nu_2>^c + \alpha_2 |\nu_1^c>) 
\end{eqnarray} 
with, $ \alpha_1 = \displaystyle \frac{m_1 m + m_2 
\tilde{m}}{m_1^2 - m_2^2} , \quad 
\alpha_2 = \displaystyle \frac{m_1 \tilde{m} + m_2 m}{m_1^2 - m_2^2} $
and $\quad {\cal N} = 2 + |\alpha_1|^2 
+ |\alpha_2|^2 $.
As mentioned earlier we now have two physical Majorana neutrino states
$|\nu^{\rm phys}_1>$ and $|\nu^{\rm phys}_2>$. The states 
$|{\nu_1^{\rm phys}}^\prime>$ and $|{\nu_2^{\rm phys}}^\prime>$ 
are related by 
$\gamma_5$ transformations to the states $|{\nu_1^{\rm phys}}>$ and 
$|{\nu_2^{\rm phys}}>$ respectively and are not independent. 

Before the electroweak phase transition, the states $|\nu_i>$ with
lepton number 1 and $|\nu_i^c>$ with a lepton number -1 were 
the physical states. They were evolving with time and their number
densities were given by the equilibrium distribution 
$n_{\nu_1} = n_{\nu_2} \sim n_\gamma$ and hence there
was no lepton asymmetry. As soon as the electroweak symmetry is
spontaneously broken, the neutrinos acquire Majorana masses and
$|\nu_1^{\rm phys}>$ and $|\nu_2^{\rm phys}>$ becomes the physical
states. The number density of $|\nu^{\rm phys}_1>$ and $|\nu^{\rm phys}_2>$
now satisfy the equilibrium distribution $n_{\nu^{\rm phys}_1} = 
n_{\nu^{\rm phys}_2} \sim
n_\gamma$. However, the lepton number of the universe is now given by,
\begin{eqnarray}
\frac{n_L}{n_\gamma} &=& \displaystyle \frac{\sum_i \left( \left| \sum_m 
<\nu^{\rm phys}_i|\nu_m> \right|^2 - \left| \sum_m 
<\nu^{\rm phys}_i|\nu^c_m> \right|^2 \right)}{
\sum_i \left( \left| \sum_m 
<\nu^{\rm phys}_i|\nu_m> \right|^2 + \left| \sum_m 
<\nu^{\rm phys}_i|\nu^c_m> \right|^2 \right)}  
\nonumber  \\
&=& \frac{g^2}{8 \pi} \frac{m_1 - m_2}{m_1 + m_2} \:\: {\rm Im}[
\sum_\alpha V_{\alpha 1}^\ast V_{\alpha 2} ]  .  \label{eqn}
\end{eqnarray}
In this  expression  $CP-$violation  comes from the $CP-$phase in
the mixing  matrix $V$.  Since we have already  made all possible
phase  rotations to work in the basis of real and  diagonal  tree
level masses for the neutrinos, the only freedom left is to phase
rotate the charged  leptons.  But then any phase  rotation of the
charged lepton fields  $l_\alpha$  keeps this Imaginary  quantity
invariant.  Unlike  the  decay   scenario considered so far in the
literature,  we  now  have  a  new
rephasing invariant  combination of the mixing matrix $(m_1 - m_2) 
{\rm Im} [\sum_\alpha V_{\alpha 1}^\ast V_{\alpha 2} ]$, which
gives rise to the $CP-$violation.  This is a
product of two matrix elements of $V$ and not four.  In the quark
sector there are no analog of this combination due to the absence
of any  Majorana  particle.  Hence  this  $CP-$violation  is also
distinct from the decay scenario.

When the doublet higgs scalar field acquires a vacuum expectation
value at a temperature of around 250 GeV, a non-zero $n_L$ as given
by the  equation(\ref{eqn})  is  spontaneously  generated  from an
initial value of $n_L = 0$.  This lepton asymmetry is same as the
$(B-L)$ asymmetry if there is no primordial  baryon asymmetry of
the  universe.  During  the  electroweak  phase  transition  this
lepton asymmetry will then be converted to a baryon asymmetry due
to the anomalous  baryon number  violation in the presence of the
sphalerons,  $$ \displaystyle \frac{n_B}{n_\gamma} 
\sim {1 \over 3} \displaystyle \frac{n_{(B-L)}}{n_\gamma}
\sim {1 \over 3}\displaystyle \frac{n_B}{n_\gamma} .  $$ 
In equation  (\ref{eqn}) there is enough freedom for us
to make the  generated  baryon  asymmetry  to be of the  order of
$O(10^{-8})$.  This is  particularly  so because  of our lack of
knowledge of the amount of $CP-$violation in the leptonic sector.

In  demonstrating  how this mechanism  works we have considered a
two generation  model and made some simplifying  assumptions.  In
the three  generation  case, this mechanism  works exactly in the
similar   way,  but  then there will be three
$CP-$violating  phases  in the  mixing  matrix  all of  which  can
contribute  to the  generation  of the lepton  asymmetry.  It is also
possible to relax the  assumption on the hierarchy of the masses,
and solve  the  problem  exactly  \cite{pas2},  which  will  only
increase our freedom to adjust the value of the generated  baryon
asymmetry.  The source of  $CP-$violation  and the  redundancy of
the out-of-equilibrium condition is not altered.

To summarize, a conceptually new model of baryogenesis has  been proposed.
When the $SU(2)_L$ higgs doublet acquires a $vev$, a
$CP-$violation  in the  Majorana  mass matrix of the  left-handed
neutrinos will make the universe lepton asymmetric spontaneously,
which then  generates a baryon  asymmetry in the  presence of the
sphaleron  fields.  Out-of-equilibrium  condition is redundant in
this scenario.

\vskip .5in

{\bf   Acknowledgement   } I  would
like to acknowledge a fellowship  from the Alexander von Humboldt
Foundation and hospitality from the Institut f\"{u}r Physik, Univ
Dortmund, Germany.

\end{document}